\documentclass[10pt]{article}
\usepackage[dvips]{epsfig}
\usepackage[T1]{fontenc}
\usepackage[latin1]{inputenc}
\usepackage{graphicx}
\usepackage[english]{babel}
\usepackage{amsmath}
\usepackage{amssymb}
\usepackage{amsfonts}
\usepackage[T1]{fontenc}
\usepackage{color}
\usepackage{babel}
\usepackage{verbatim}
\usepackage[unicode=true,pdfusetitle,bookmarks=true,bookmarksnumbered=false,bookmarksopen=false,
 breaklinks=false,pdfborder={0 0 1},backref=false,colorlinks=true]{hyperref}
\hypersetup{linkcolor=blue,citecolor=blue}
\makeatletter
\usepackage[dvips]{epsfig}
\usepackage[T1]{fontenc}
\usepackage{color}
\usepackage{pdflscape}

\begin{document}

\title{\bf FLRW-Cosmology in Generic Gravity Theories}

\author{Metin G{\" u}rses\thanks{%
email: gurses@fen.bilkent.edu.tr},~and~Yaghoub Heydarzade\thanks{%
email: yheydarzade@bilkent.edu.tr}
\\{\small Department of Mathematics, Faculty of Sciences, Bilkent University, 06800 Ankara, Turkey}}

\date{\today}

\maketitle
\begin{abstract}
We prove that for the Friedmann-Lemaitre-Robertson-Walker
metric, the field equations of  any generic gravity theory in arbitrary dimensions are of the perfect fluid type. The cases of general Lovelock and $\mathcal{F}(R, \mathcal{G})$ theories are given as examples.
\end{abstract}
\maketitle
\section{Introduction}
The Friedmann-Lemaitre-Robertson-Walker (FLRW) metric is the most known and most studied metric in General Relativity
(GR). FLRW metric is mainly used to describe the universe  as a homogeneous isotropic fluid distribution \cite{books,
peeb, islam, ellis}. For inhomogeneous
cosmological solutions, see for example \cite{kras, us}. On the other hand, current cosmological observations
indicate that our universe is undergoing an accelerating expansion phase. The origin of this accelerating expansion still remains an
open question in cosmology. Several approaches  for explaining the current
accelerated
expanding phase have
been proposed in the literature such as introducing cosmological constant
\cite{paddy}, dynamical
dark energy models and modified theories of gravity \cite{mt, noj, faro,
cope}.
Amongst the latter, higher order curvature corrections to Einstein's field equations have been considered by several authors \cite{odint, love, casal,
jose}.
In the context of modified theories, some attempts for a geometric interpretation
of the dark side of the universe as a perfect fluid have been done  \cite{capoz1, capoz2, capoz3, odin1, odin2, odin3} but the picture is not complete yet.
 In this work, we put one step forward to prove that the perfect fluid from
of the dark component of the Universe is true for any generic modified theory of
gravity. A generic gravity theory derivable from a variational principle
can be given by the action

\begin{equation}
I=\int\, d^{D}\,x\, \sqrt{-g}\,\left (\frac{1}{\kappa} \left(R-2\Lambda \right)+
\mathcal{F}(g,\,\mbox{Riem},\,\nabla \mbox{Riem},\, \nabla \nabla \mbox{Riem}, \cdots)+ \mathcal{L}_{M} \right),
\end{equation}
where $g$, $\mbox{Riem}$, $\nabla \mbox{Riem}$, $\nabla \nabla \mbox{Riem}$, etc in $\mathcal{F}$ denote  the spacetime metric, Riemann tensor and its covariant derivatives at any order, respectively, and $\mathcal{L}_M$ is the Lagrangian of the matter
fields.  The function $\mathcal{F}(g,\,\mbox{Riem},\,\nabla \mbox{Riem},\, \nabla \nabla \mbox{Riem}, \cdots) $ is the part of the Lagrange function corresponding to higher order couplings, constructed from the metric, the Riemann tensor and its covariant derivatives. The corresponding field equations are
\begin{equation}
\frac{1}{\kappa}\left(G_{\mu\nu}  +\Lambda g_{\mu\nu} \right)+ \mathcal{E}_{\mu \nu}=T_{\mu\nu}.
\end{equation}
Here $\mathcal{E}_{\mu \nu}$  is a symmetric divergent free tensor obtained from the variation of $\mathcal{F}(g,\, \mbox{Riem},\,\nabla \mbox{Riem},\, \nabla \nabla \mbox{Riem}, \cdots)$ with respect to the spacetime metric $g_{\mu\nu}$. Our treatment, in this work, is to consider this tensor, $\mathcal{E}_{\mu \nu}$, as any second rank tensor obtained from the Riemann tensor and its covariant derivatives at any order. Since the Ricci tensor $R_{\mu \nu}$ and Ricci scalar $R$ are obtainable from the Riemann tensor we did not consider the function $\mathcal{F}$ depending on explicitly on the Ricci tensor and Ricci scalar. There are some works  showed recently that the tensor $\mathcal{E}_{\mu \nu}$ takes the perfect fluid form for the FLRW spacetimes when the function $\mathcal{F}$ depends only the Ricci and the Gauss-Bonnet scalars $R$ and $\mathcal{G}$ respectively \cite{capoz1,
capoz2}, as well as  the  Ricci scalar $R$ and $\square R$ of any order \cite{capoz3}. In the present work, we prove that the tensor $\mathcal{E}_{\mu \nu}$ takes the perfect fluid form for any generic modified gravity theory in the FLRW spacetimes in arbitrary dimensions. We then apply our result to two special cases $\mathcal{F}(R,\mathcal{G})$  and Lovelock theory in any dimension $D$.

The organization of the paper is as follows. In Section 2, we give the covariant description of D-dimensional FLRW metric and derive all the corresponding
geometrical quantities. In section 3, we introduce the closed FLRW-tensor algebra by proving that all the geometrical quantities for FLRW spacetimes, the curvature tensor and it's covariant derivatives at any order, are expressed in terms of the metric tensor $g_{\mu\nu}$ and the product $u_{\mu} u_{\nu}$ where $u_{\mu}$ is the unit timelike tangent vector of the timelike geodesic. By using this property, i.e., the existence of a closed tensor algebra, we prove a theorem on the field equations of  generic gravity theories. In Sections 4 and 5,  we use the proved theorem to write the field equations of Lovelock and $\mathcal{F}(R, \mathcal{G})$ theories, respectively. Section 6 is devoted to our concluding remarks.

\section{Covariant Description of the FLRW Spacetimes in D-Dimensions}

We begin with the definition of the $D$-dimensional FLRW spacetimes.

\vspace{0.4cm}
\noindent
{\bf Definition 1}: {\it  The $D$-dimensional FLRW spacetime is defined with the following metric

\begin{equation}\label{metric}
g_{\mu\nu}=-u_\mu\, u_\nu+a^2\,h_{\mu\nu},
\end{equation}
where $x^{\mu}=(t, x^{i})$, $\mu,\,\nu=0,...,D-1$, $a=a(t),~u_\mu=\delta^0_\mu$, and $h_{\mu\nu}$
reads as
\begin{equation}
h_{\mu\nu}=\begin{pmatrix}
0 & 0 & \hdots & 0 \\
0 &  &  &  \\
\vdots &  & h_{ij} &  \\
0 &  &  &  \\
\end{pmatrix},
\end{equation}
where $h_{ij}=h_{ij}(x^{a})$  with $i,j=1,...,D-1$ is the metric of a space of constant curvature $k$.}

\vspace{0.4cm}
\noindent
One can verify\begin{eqnarray}
&& u^\mu \,h_{\mu\nu}=u_\mu\, h^{\mu\nu}=0,\nonumber\\
&&h^{\mu}_{\alpha}= h^{\mu\alpha}\,h_{\alpha\nu}=\delta^{\mu}_{\nu}
+u^\mu \,u_\nu.
\end{eqnarray}
The corresponding Christoffel symbols to the metric (\ref{metric}) can be obtained as
\begin{equation}\label{christ}
\Gamma^\mu_{\alpha \beta}=\gamma^\mu_{\alpha\beta}-a\,\dot a\, u^\mu\, h_{\alpha \beta}+H\,\left( 2u_\alpha\, u^\mu\, u_\beta +u_\beta\, \delta^\mu_\alpha +u_\alpha\, \delta^\mu_\beta \right),
\end{equation}
where the dot sign represents the derivative with respect to time $t$, $H=\dot a/a$ is the Hubble parameter and $\gamma^\mu_{\alpha\beta}$ is defined as
\begin{equation}
\gamma^\mu_{\alpha\beta}=\frac{1}{2}a^2\, h^{\mu\gamma}\,\left(h_{\gamma \alpha,\beta}+h_{\gamma \beta,\alpha} -h_{\alpha \beta,\gamma} \right).
\end{equation}
One can also prove the following properties for $u_\alpha$ and $h_{\alpha\beta}$
\begin{eqnarray}
&&u_\mu\,h^\mu_{\alpha\gamma,\beta}=0= u_\mu\,\gamma^\mu_{\alpha\beta},\nonumber\\
&&\nabla_{\alpha}u_\beta=-a\,\dot a \,h_{\alpha\beta}=-H\left(g_{\alpha \beta}+u_\alpha\,
u_\beta  \right),\label{u}\nonumber\\
&&\nabla_{\gamma}h_{\alpha\beta}=-H\left(2u_\gamma\, h_{\alpha\beta}+u_\beta\, h_{\gamma\alpha}  +u_\alpha\, h_{\gamma\beta}    \right)=-\frac{\dot a}{a^3}\left(2u_\gamma \,g_{\alpha\beta}+u_\beta\, g_{\gamma\alpha}  +u_\alpha\, g_{\gamma\beta}  +4u_\alpha\, u_\beta\, u_\gamma \right).
\end{eqnarray}
Using the Christoffel symbols (\ref{christ}), one can find the components
of the  Riemann curvature tensor as
\begin{eqnarray}\label{riemann}
R^{\mu}_{\alpha\beta\gamma}&=&\partial_\beta\Gamma^\mu_{\alpha\gamma}-\partial_\gamma\Gamma^\mu_{\beta\alpha}+\Gamma^\mu_{\beta\rho}\Gamma^\rho_{\alpha\gamma}-\Gamma^\mu_{\gamma\rho}\Gamma^\rho_{\beta\alpha}\nonumber\\
&=&r^{\mu}_{\alpha\beta\gamma}-\dot H\, u_\alpha\left(u_\gamma\,\delta^\mu_\beta
-u_\beta\,\delta^\mu_\gamma \right)+\left( \dot a^2+a\ddot a\right)u^\mu\left( u_\gamma\, h_{\alpha\beta} -u_\beta\, h_{\alpha\gamma} \right)\nonumber\\
&&+H^2\left( u_\beta\, u_\alpha\,\delta^\mu_\gamma
-u_\gamma\, u_\alpha\,\delta^\mu_\beta  \right)\nonumber\\
&&-\,\dot a^2\left(\delta^\mu_\beta\, h_{\alpha\gamma}+\delta^\mu_\gamma \,h_{\alpha\beta}
 -2u^\mu\, u_\beta\, h_{\alpha\gamma}+2u^\mu\, u_\gamma\, h_{\alpha\beta}\right),
\end{eqnarray}
where the curvature tensor $r^{\mu}_{\alpha\beta\gamma}$ is defined as
\begin{equation}
r^{\mu}_{\alpha\beta\gamma}=\gamma^{\mu}_{\alpha\gamma,\beta}-\gamma^{\mu}_{\alpha\beta,\gamma}+
\gamma^{\mu}_{\beta\rho}\,\gamma^{\rho}_{\alpha\gamma}-\gamma^{\mu}_{\gamma\rho}\,\gamma^{\rho}_{\alpha\beta}\, .
\end{equation}
On the other hand, the curvature tensor   $r^{\mu}_{\alpha\beta\gamma}$ for a
 Riemannian space with the constant curvature $k$ can be written
as
\begin{eqnarray}\label{riem}
r^{\mu}_{\alpha\beta\gamma}=k\left( h^\mu_\beta\, h_{\alpha\gamma}-h^\mu_\gamma\, h_{\alpha\beta}   \right),
\end{eqnarray}
where it vanishes if one of  $\mu,\nu,\alpha$ or $\gamma$ is zero.
\\
Using (\ref{metric}) and (\ref{riem}), the components
of the  Riemann curvature tensor (\ref{riemann}) can be written in the following
linear form in terms of the metric  $g_{\mu\nu}$ and the four vector
$u_\mu$
\begin{equation}\label{riemann2}
R_{\mu \alpha\beta\gamma}=\left(  g_{\mu_\beta} g_{\alpha\gamma}-g_{\mu_\gamma} g_{\alpha\beta}\right)\rho_1+\left(u_{\mu}\left(g_{\alpha\gamma} u_\beta -  g_{\alpha\beta} u_\gamma \right)-u_\alpha\,\left(g_{\mu_\gamma} u_\beta -  g_{\mu_\beta} u_\gamma \right)\right)\rho_2,
\end{equation}
where $\rho_1$ and $\rho_2$ are defined as
\begin{eqnarray}
&&\rho_1=H^2 +\frac{k}{a^2},\label{rho1}\\
&&\rho_2=H^2 +\frac{k}{a^2}-\frac{\ddot a}{a}=-\dot{H}+\frac{k}{a^2}. \label{rho2}
\end{eqnarray}
 The contractions of the Riemann tensor (\ref{riemann2}) gives the Ricci tensor
and Ricci scalar, respectively, as
\begin{eqnarray}\label{ricci}
&&R_{\alpha\gamma}=g_{\alpha\gamma}\left( (D-1)\rho_1-\rho_2 \right)+u_\alpha u_\gamma(D-2)\rho_2, \nonumber\\
&&R=\left(D-1 \right)\left(D\rho_1 -2\rho_2 \right).
\end{eqnarray}
One can also verify that the Weyl tensor
defined as\begin{equation}\label{weyl}
C^{\mu}_{\alpha\beta\gamma}=R^{\mu}_{\alpha\beta\gamma}+\frac{1}{D-2}\left( \delta^\mu_\gamma R_{\alpha\beta}-\delta^\mu_\beta R_{\alpha\gamma}+ g_{\alpha\beta}R^\mu_\gamma - g_{\alpha\gamma} R^\mu_\beta\right)+\frac{1}{(D-1)(D-2)}\left(  \delta^\mu_\beta g_{\alpha\gamma}-\delta^\mu_\gamma g_{\alpha\beta}\right)R,
\end{equation}
vanishes for the metric (\ref{metric}). Hence we have
the following theorem \cite{chen,Man1,Man2, Man3}:

\vspace{0.5cm}
\noindent
{\bf Theorem 2:}\, {\it FLRW spacetimes are conformally flat for all values of spatial curvature
$k$ in any dimensions.}

\section{FLRW-Tensor Algebra}

For some spacetimes, such as spherically symmetric  and Kerr-Schild-Kundt spacetimes, it is possible to simplify the field equations of any generic gravity theories. To achieve such a simplification we need a closed tensorial algebra. By the of use this tensorial algebra, the goal is to find the most general symmetric and second rank tensor in this tensor algebra . This is the way of finding universal metrics in general relativity \cite{gur-ser, gur-sis-tekin1, gur-sis-tekin2}. In this section, we construct such a closed tensor algebra for the $D$-dimensional FLRW spacetimes and with the use of this tensor algebra we show that the field equations of any generic gravity theory, in $D$-dimensional FLRW spacetimes, have the perfect fluid form.

The geometrical tensors, Riemann and Ricci, are expressed solely by the metric tensor  $g_{\mu\nu}$ and the timelike vector $u_{\mu}$ as
\begin{eqnarray}
&&R_{\mu\alpha\beta\gamma}=\left(  g_{\mu\beta} g_{\alpha\gamma}-g_{\mu\gamma} g_{\alpha\beta}\right)\rho_1+\left(u_{\mu}\left(g_{\alpha\gamma} u_\beta -  g_{\alpha\beta} u_\gamma \right)-u_\alpha\,\left(g_{\mu\gamma} u_\beta -  g_{\mu\beta} u_\gamma \right)\right)\rho_2,\nonumber\\
&&R_{\alpha\gamma}=g_{\alpha\gamma}\left( (D-1)\rho_1-\rho_2 \right)+u_\alpha u_\gamma(D-2)\rho_2, \nonumber\\
&&R=\left(D-1 \right)\left(D\rho_1 -2\rho_2 \right),
\end{eqnarray}
where $\rho_{1}$ and $\rho_{2}$ are defined in (\ref{rho1}) and (\ref{rho2}), respectively.
Not only these tensors but also tensors produced by taking the covariant derivatives of them are also represented by the metric tensor $g_{\alpha\beta}$ and the vector $u_{\alpha}$. As examples, the covariant derivatives of the four
vector $u_\alpha$ and the Ricci tensor $R_{\alpha\beta}$ are given as follows
\begin{eqnarray}
&&\nabla_{\alpha}u_\beta=-H\left(g_{\alpha \beta}+u_\alpha\,
u_\beta  \right),\nonumber\\
&&\nabla_{\gamma}\,R_{\alpha \beta}=[(D-2) \dot{\rho}_{1}-\dot{\rho}_{2}]\, g_{\alpha \beta}\, u_{\gamma}
-(D-2) \rho_{2} \dot{H} (g_{\alpha \gamma} u_{\beta}
+g_{\beta \gamma} u_{\alpha})-2(D-2) \rho_{2} \dot{H} u_{\alpha} u_{\beta} u_{\gamma},
\end{eqnarray}
and consequently one can obtain
\begin{eqnarray}
&&{\square R}_{\alpha \beta}=-[\ddot P+(D-1) H \dot P-2 Q H^2]\, g_{\alpha\beta}+[2 D Q H^2-\ddot Q+ 2(D-1) H \dot Q] u_{\alpha} u_{\beta},\nonumber\\
&&\square R=-D \ddot P-D(D-1) H \dot P+\dot Q-2(D-1) H \dot Q,
\end{eqnarray}
where $P$ and $Q$ are defined as
\begin{eqnarray}
&&P=(D-1) \rho_{1}-\rho_{2},\nonumber\\
&&Q=(D-2) \rho_{2}.
\end{eqnarray}
The covariant derivative of the Riemann tensor has the similar structure. We have the similar structure for the higher order covariant derivatives of the Riemann and Ricci tensors. They are all expressed as the sum of monomials  of the same rank which are products of the metric tensor $g_{\mu\nu}$ and the vector $u_{\mu}$.

\vspace{0.4cm}
\noindent
{\bf Definition 3}: {\it A tensor $M$ of rank $k$ denoting the monomials of the product of metric and the vector $u_{\mu}$ is given by

\begin{equation}
M_{\mu_{1} \mu_{2} \mu_{3} \mu_{4} \cdots  \mu_{k}}=g_{\mu_{1} \mu_{2}} g_{\mu_{3} \mu_{4}} \cdots u_{\mu_{k-1}} u_{\mu_{k}}
\end{equation}
There are $r$ number of metric tensor and $k-r$ number of vector $u_{\mu}$ in a monomial of rank $k$. Here $r$ is any nonnegative integer.
}

\vspace{0.4cm}
\noindent
{\bf Proposition 4}: {\it  In $D$-dimensional FLRW spacetimes any tensor generated by the curvature tensor and its covariant derivatives at any order is the sum of the different monomials of the same rank.}

\vspace{0.4cm}
\noindent
All scalars and functions depend only on the time variable $t$. Hence, the
derivative of the Ricci scalar  is given by
\begin{equation}
\nabla_{\gamma} R= \dot{R} \, u_{\gamma}.
\end{equation}
This is valid also for any scalars obtained from the Riemann and Ricci tensors and their covariant derivatives at any order. Let $\Theta$ be any of such a scalar then

\begin{equation}
\nabla_{\gamma} \Theta= \dot{\Theta} \, u_{\gamma}.
\end{equation}

Now we are ready to obtain the most general symmetric and second rank tensor from the contractions of higher order tensors. For illustration, let us consider the following example.
If  $E_{\alpha_{1} \alpha_{2} \cdots \alpha_{m}}$ is a tensor of
rank $m$ obtained from the Ricci and Riemann tensors and their covariant derivatives at any order, then, by Proposition 4,  it takes the following form for $m= \mbox{even integer}$
\begin{eqnarray}
E_{\alpha_{1} \alpha_{2} \cdots \alpha_{m}}&=& A_{1}\, g_{\alpha_{1} \alpha_{2}} \cdots g_{\alpha_{m-1} \alpha_{m}} +A_{2}\, g_{\alpha_{1} \alpha_{2}} \cdots u_{m-1} u_{\alpha_{m}}+\cdots  \nonumber\\
&&+A_{m-1}\, g_{\alpha_{1} \alpha_{2}}\, u_{\alpha_{3}} \cdots u_{m}+A_{m}\, u_{\alpha_{1}}\, u_{\alpha_{2}} \cdots u_{\alpha_{m}},
\end{eqnarray}
and for $m=\mbox{odd integer}$ as
\begin{eqnarray}
E_{\alpha_{1} \alpha_{2} \cdots \alpha_{m}}&=& B_{1}\, g_{\alpha_{1} \alpha_{2}} \cdots g_{\alpha_{m-2} \alpha_{m-1}}\,u_{\alpha_{m}} +B_{2}\, g_{\alpha_{1} \alpha_{2}} \cdots u_{\alpha_{m-2}}\, u_{\alpha_{m-1}} u_{\alpha_{m}}+\cdots  \nonumber\\
&&+B_{m-1}\, g_{\alpha_{1} \alpha_{2}}\, u_{\alpha_{3}} \cdots u_{m}+B_{m}\, u_{\alpha_{1}}\, u_{\alpha_{2}} \cdots u_{\alpha_{m}},
\end{eqnarray}
where $A_{k},B_{k}$ ($k=1,2, \cdots , m$) are functions of the time parameter
$t$. All the tensors of rank two obtained by the contraction of such tensors are of our interests. To see the result of such a contraction, let us consider the contraction of the monomials  of the metric tensor $g_{\mu\nu}$ and the vector $u_{\mu}$. As an example

\begin{equation}
g_{\alpha_{1} \alpha_{2}} \, g_{\alpha_{3} \alpha_{4}}\,u_{\alpha_{5}}\, u_{\alpha_{6}} \, u_{\alpha_{7}},
\end{equation}
is a monomial of rank seven. Since $u_{\alpha}\, u^{\alpha}=-1$ and $g_{\mu\nu}$ is the metric tensor then any second rank tensor obtained from the contraction of such two different monomials is either
$g_{\mu \nu}$ or   $u_{\mu} u_{\nu}$. Therefore, if $E_{\mu \alpha_{1} \alpha_{2} \cdots \alpha_{m}}$ and $F_{\nu}\,^{\alpha_{1} \alpha_{2} \cdots \alpha_{m}}$ are two tensors obtained from the Riemann, Ricci tensors and their covariant derivatives at any order, then
we have
\begin{equation}
 E_{\mu \alpha_{1} \alpha_{2} \cdots \alpha_{m}}\,F_{\nu}\,^{\alpha_{1} \alpha_{2} \cdots \alpha_{m}}=C_{1}\, g_{\mu \nu}+C_{2} u_{\mu} u_{\nu}, \label{ozel}
\end{equation}
 where $C_{1}$ and $C_{2}$ are some scalars. In the general case the idea of obtaining a symmetric and second rank tensor from the above tensor algebra is similar. The main points are: (1) all tensors are the sum of monomials of the metric tensor and the vector $u_{\mu}$, (2) any symmetric tensor of the second rank obtained from the products of monomials is either the metric tensor $g_{\mu \nu}$ or $u_{\mu}\, u_{\nu}$, and (3) due to the first two facts any symmetric second rank tensor obtained from the curvature tensor and its covariant derivatives at any order will be similar to (\ref{ozel}).  Then, we have the following theorem:

\vspace{0.3cm}
\noindent
{\bf Theorem 5}: {\it Any second rank tensor obtained from the metric tensor, Riemann tensor,  Ricci tensor, scalar $\psi$ and their covariant derivatives at any order is a combination of the metric tensor $g_{\mu \nu}$ and $u_{\mu} u_{\nu}$ that is
\begin{equation}\label{E}
\mathcal{E}_{\mu \nu}=A g_{\mu \nu}+B u_{\mu} u_{\nu},
\end{equation}
where $A$ and $B$ are functions of $a(t)$ and $\psi(t)$ and their time derivatives at any order.}

\vspace{0.5cm}
\noindent
Some special cases of this theorem are given in \cite{capoz1, capoz2, capoz3}. In these references, this theorem was proved for the field equations of special cases $\mathcal{F}(R,\mathcal{G})$ and $\mathcal{F}(R,\, \square R,\,\square \square R, \cdots)$. In  \cite{capoz3}, the considered geometry  is the generalized FLRW spacetime.
We have the following corollary of this theorem:

\vspace{0.3cm}
\noindent
{\bf Corollary 6}: {\it The field equations of any generic gravity theory takes the form

\begin{equation}\label{FEs}
G_{\mu \nu}+ \Lambda g_{\mu \nu}+\mathcal{E}_{\mu \nu}=T_{\mu \nu},
\end{equation}
where $G_{\mu \nu}$ is  the Einstein tensor, $\Lambda$ is the cosmological constant, $T_{\mu \nu}$ is the energy momentum tensor of perfect fluid distribution and $\mathcal{E}_{\mu \nu}$ comes from the higher order curvature terms. Hence the general field equations take the form
\begin{eqnarray}\label{rp}
&&\rho=\frac{1}{2}(D-1)(D-2) \rho_{1}-\Lambda +B-A,\nonumber\\
&&p=(D-2)\left[-\frac{1}{2}(D-1) \rho_{1}+\rho_{2}\right] +\Lambda+A.
\end{eqnarray}}
Thus,  regarding (\ref{rp}), the interpretation of  $A$ and $B$ in $\mathcal{E}_{\mu \nu}$ tensor (\ref{E}) is as follows. $A$ is the effective pressure, and the
combination $B-A$ is the sum of effective pressure and effective energy density of an effective perfect fluid of the geometric
origin.   As the applications of the theorem in the following sections, we prove that the field equations of the Einstein-Lovelock
theory and a generalized version of  Einstein-Gauss-Bonnet  theory $\mathcal{F}(R, \mathcal{G})$, as two examples for general higher order curvature
theories, reduce to the perfect fluid form with the energy density $\rho$ and pressure $p$ given in  (\ref{rp}).
%%%%%%%%%%%%%%%%%%%%%%%%%%%%%%%%%%%%%%%%%%%%%%%%%%%%%%%%%%%%%%%%%%%%%%%%%%
\section{Einstein-Lovelock Theory }

The action of the Lovelock theory in $D$-dimensions is given by \cite{love}

\begin{equation}
I=\int\, d^{D}\,x\, \sqrt{-g}\,\left (\frac{1}{\kappa} \left(R-2\Lambda \right)+\sum_{n=2}^{N}\,\alpha_{n}\, L_{n}\right),
\end{equation}
where $\alpha_{n}$'s are constants and
\begin{equation}
L_{n}=2^{-n}\, \delta^{\mu_{1} \mu_{2} \cdots \mu_{2n}}\,_{\nu_{1} \nu_{2} \cdots \nu_{2n}} R^{\nu_{1} \nu_{2}}\,_{\mu_{1} \mu_{2}} \,
R^{\nu_{3} \nu_{4}}\,_{\mu_{3} \mu_{4}} \cdots R^{\nu_{2n-1} \nu_{2n}}\,_{\mu_{2n-1} \mu_{2n}}.
\end{equation}
The corresponding field equations take the form \cite{love}

\begin{equation}\label{gb1}
\frac{1}{\kappa}\left(G_{\mu\nu}  +\Lambda g_{\mu\nu} \right)+ \sum_{n=2}^{N}\alpha_{n}\,(\mathcal{H}_{\mu\nu})_{n}=T_{\mu\nu},
\end{equation}
where  the tensor $(\mathcal{H}_{\mu\nu})_{n}$ is given by \cite{casal}
\begin{equation}\label{htensor}
(\mathcal{H}^{\mu}\,_{\nu})_{n}=\frac{1}{2^{n+1}} \delta^{\mu \alpha \beta \alpha_{1} \beta_{1} \cdots \alpha_{n} \beta_{n}}_{\nu \gamma \sigma \gamma_{1} \sigma_{1} \cdots \gamma_{n} \sigma_{n}}\, R_{\alpha \beta}\,^{\gamma \sigma}\, R_{\alpha_{1} \beta_{1}}\,^{\gamma_{1} \sigma_{1}}\, \cdots R_{\alpha_{n} \beta_{n}}\,^{\gamma_{n} \sigma_{n}}.
\end{equation}
In the case of the FLRW metric, $(\mathcal{H}_{\mu\nu})_{n}$  reduces to
the following form
 \begin{equation}
({\mathcal H}_{\mu \nu})_{n}=\frac{n\,(D-2)!}{(D-2n-1)!}\,( \rho_{1})^{n-1}\,\left[(\rho_{2}-\frac{D-1}{2\,n} \rho_{1}) g_{\mu \nu}+\rho_{2} u_{\mu} u_{\nu} \right],
\end{equation}
representing a linear combination of metric $g_{\mu\nu}$ and $u_\mu u_\nu$. Then, we have the following proposition.
\\
\\
\vspace{0.5cm}
\noindent
{\bf Proposition 7:}
{\it The pressure $p$ and the energy density $\rho$ in the
context of Einstein-Lovelock theory for any $n$ can be obtained as
\begin{eqnarray}\label{ener}
&&p=\frac{1}{\kappa}\left[\left(D-2\right)\left(\rho_2-\frac{1}{2}(D-1)\rho_1 \right)+\Lambda \right]+\sum_{n=2}^N \alpha_n\frac{n\,(D-2)!}{(D-2n-1)!}\,( \rho_{1})^{n-1}\left(\rho_{2}-\frac{D-1}{2\,n} \rho_{1}\right), \nonumber\\
&&\rho=\frac{1}{\kappa}\left[\frac{(D-1)\left(D-2\right)}{2}\rho_1 -\Lambda \right]+\sum_{n=2}^N \alpha_n\frac{\,(D-1)!}{2(D-2n-1)!}\,( \rho_{1})^{n}. \end{eqnarray}}
When $k=0$ and a barotropic equation of state  $p=w \rho$ is considered,
the Hubble parameter  $H$ satisfies the following first order nonlinear ordinary differential equation
\begin{eqnarray}
&&\left[\frac{(D-2)}{2\kappa}+\sum_{n=2}^N n \,\,\bar{\alpha}_n \,\,H^{2n-2} \right]\, \dot H= \nonumber\\
&&-(w+1)\, \left(\frac{1}{\kappa}\left[\frac{(D-1)\left(D-2\right)}{2}\,H^2 -\Lambda \right]+\frac{1}{2}\,\sum_{n=2}^N (D-1)\,\bar{\alpha}_n \, H^{2n}\right), \label{firstorder}
\end{eqnarray}
where
\begin{equation}
\bar{\alpha}_{n}=\frac{(D-2)!}{(D-2n-1)!}\alpha_n,
\end{equation}
are the re-scaled coupling constants of the theory. The case $H=\mbox{constant}$
solves the equation (\ref{firstorder}) for all $D$ and $n$ but the energy density $\rho$ vanishes for this kind of solutions with a linear equation of state. For any $D$ and $n$ it is possible to integrate the above equation (\ref{firstorder}) and the solution is given in the following proposition.

\vspace{0.5cm}
\noindent
{\bf Proposition 8}: {\it Let the polynomial
\begin{equation}
P_{N}(H^2)=\frac{1}{\kappa}\left[\frac{(D-1)\left(D-2\right)}{2}\,H^2 -\Lambda \right]+\frac{1}{2}\,\sum_{n=2}^N (D-1)\,\bar{\alpha}_n \, H^{2n},
\end{equation}
of $H^2$ and of the degree $N$ has the $N$ roots  $k_{i}^2$ $(i-1,2,\cdots ,N)$, then the solution of the equation (\ref{firstorder}) is given by
\begin{equation}
\sum_{n=1}^{N}\, p_{n} \tanh^{-1} \left(q_{n}\,\frac{H}{k_{n}} \right)=t-t_{0},
\end{equation}
where $p_{i}$ and $q_{i}$ are some constants depending on the constants of the theory.}

\vspace{0.5cm}
\noindent
The exact solutions corresponding to $n=2$ and as $N \to \infty $ will be
discussed in \cite{gur=hey}.

\section{ Generalized Einstein-Gauss-Bonnet Theory }
The generalization of the action of the Einstein-Gauss-Bonnet theory  is given by

\begin{equation}
I=\int\, d^Dx\, \sqrt{-g} \left(\frac{1}{\kappa}\left(R -2\Lambda\right)+\alpha \mathcal{F}(R, \mathcal{G})\right)+\int\, d^Dx\, \sqrt{-g}\,\mathcal{L}_M,
\end{equation}
where $\mathcal{G}$ represents the Gauss-Bonnet topological invariant, i.e $\mathcal{G}=R_{\alpha \beta\rho\sigma}\,R^{\alpha \beta \rho\sigma}-4R_{\alpha\beta }R^{\alpha\beta}+R^2$.
The corresponding field equations read as
\begin{equation}\label{gb}
\frac{1}{\kappa}\left(G_{\alpha\beta}+\Lambda g_{\alpha\beta} \right)+ \alpha \mathcal{E}_{\alpha\beta}=T_{\alpha\beta},
\end{equation}
where the modified Einstein-Gauss-Bonnet  tensor $\mathcal{E}_{\alpha\beta}$ is given by
\begin{eqnarray}\label{E1}
\mathcal{E}_{\alpha\beta}&=&-\frac{1}{2} \mathcal{F}(R, \mathcal{G})\, g_{\alpha \beta}\, +\mathcal{F}_{R}(R, \mathcal{G})\, R_{\alpha \beta}-\nabla_{\alpha} \nabla_{\beta} \mathcal{F}_{R}(R, \mathcal{G})+
 g_{\alpha \beta}\nabla^2 \mathcal{F}_{R}(R, \mathcal{G})\nonumber\\
 &&+2\left(R\,R_{\alpha\beta}-2{R^\rho}_\alpha\,R_{\beta\rho}+2R_{\alpha\rho\sigma\beta}\,R^{\rho\sigma}+R_{\beta\mu \nu\gamma}\,{R_{\alpha}}^{\mu\nu\gamma}\right)\mathcal{F}_{\mathcal{G}}(R, \mathcal{G})\nonumber\\
&&-2R\left( \nabla_\alpha\nabla_\beta-g_{\alpha\beta}\nabla^2\right)\mathcal{F}_{\mathcal{G}}(R, \mathcal{G})+4\left({R^\mu}_\beta \nabla_\mu\nabla_\alpha+{R^\mu}_\alpha \nabla_\mu\nabla_\beta \right)\mathcal{F}_{\mathcal{G}}(R, \mathcal{G})\nonumber\\
&&-4\left(R_{\alpha\beta} \nabla^2+g_{\alpha\beta}R^{\mu\nu} \nabla_\mu\nabla_\nu
+ R_{\alpha\rho\sigma\beta}\nabla^\rho\nabla^\sigma\right)\mathcal{F}_{\mathcal{G}}(R, \mathcal{G}),
\end{eqnarray}
where $\mathcal{F}_{R}=\frac{\partial \mathcal{F}}{\partial R}$ and $\mathcal{F}_{\mathcal{G}}=\frac{\partial \mathcal{F}}{\partial \mathcal{G}}$.

One can define a second rank tensor $H_{\alpha\beta}$ as

\begin{equation}\label{h}
H_{\alpha \beta}=2\left[R\,R_{\alpha\beta}-2{R^\rho}_\alpha\,R_{\beta\rho}+2R_{\alpha\rho\sigma\beta}\,R^{\rho\sigma}+R_{\beta\mu \nu\gamma}\,{R_{\alpha}}^{\mu\nu\gamma}-\frac{1}{4} \mathcal{G}\, g_{\alpha\beta}\right],
\end{equation}
which vanishes  in four dimensions \cite{met}. Then, $\mathcal{E}_{\alpha\beta}$
can be written in terms of the $H_{\alpha \beta}$ as
\begin{eqnarray}\label{E10}
\mathcal{E}_{\alpha\beta}&=&-\frac{1}{2} \mathcal{F}(R, \mathcal{G})\, g_{\alpha \beta}\, +\mathcal{F}_{R}(R, \mathcal{G})\, R_{\alpha \beta}-\nabla_{\alpha} \nabla_{\beta} \mathcal{F}_{R}(R, \mathcal{G})+
 g_{\alpha \beta}\nabla^2 \mathcal{F}_{R}(R, \mathcal{G})\nonumber\\
 &&+\left(H_{\alpha \beta}+\frac{1}{2}\mathcal{G}g_{\alpha\beta}\right)\mathcal{F}_{\mathcal{G}}(R, \mathcal{G})\nonumber\\
&&-2R\left( \nabla_\alpha\nabla_\beta-g_{\alpha\beta}\nabla^2\right)\mathcal{F}_{\mathcal{G}}(R, \mathcal{G})+4\left({R^\mu}_\beta \nabla_\mu\nabla_\alpha+{R^\mu}_\alpha \nabla_\mu\nabla_\beta \right)\mathcal{F}_{\mathcal{G}}(R, \mathcal{G})\nonumber\\
&&-4\left(R_{\alpha\beta} \nabla^2+g_{\alpha\beta}R^{\mu\nu} \nabla_\mu\nabla_\nu
+ R_{\alpha\rho\sigma\beta}\nabla^\rho\nabla^\sigma\right)\mathcal{F}_{\mathcal{G}}(R, \mathcal{G}).
\end{eqnarray}

 Hence,   in four dimensions, $\mathcal{E}_{\alpha\beta}$ (\ref{E1}) reduces to the following form
\begin{eqnarray}\label{eab}
\mathcal{E}_{\alpha\beta}&=&-\frac{1}{2} \mathcal{F}(R, \mathcal{G})\, g_{\alpha \beta}\, +\mathcal{F}_{R}(R, \mathcal{G})\, R_{\alpha \beta}-\nabla_{\alpha} \nabla_{\beta} \mathcal{F}_{R}(R, \mathcal{G})+g_{\alpha \beta}
 \nabla^2 \mathcal{F}_{R}(R, \mathcal{G}) \nonumber\\
 &&+\frac{1}{2}\mathcal{G}\mathcal{F}_{\mathcal{G}}(R, \mathcal{G})-2R\left( \nabla_\alpha\nabla_\beta-g_{\alpha\beta}\nabla^2\right)\mathcal{F}_{\mathcal{G}}(R, \mathcal{G})+4\left({R^\mu}_\beta \nabla_\mu\nabla_\alpha+{R^\mu}_\alpha \nabla_\mu\nabla_\beta \right)\mathcal{F}_{\mathcal{G}}(R, \mathcal{G})\nonumber\\
&&-4\left(R_{\alpha\beta} \nabla^2+g_{\alpha\beta}R^{\mu\nu} \nabla_\mu\nabla_\nu
+ R_{\alpha\rho\sigma\beta}\nabla^\rho\nabla^\sigma\right)\mathcal{F}_{\mathcal{G}}(R, \mathcal{G}).
\end{eqnarray}
The geometric tensor $\mathcal{E}_{\alpha\beta}$ (\ref{eab}) corresponds to the tensor
 $\Sigma_{\alpha\beta} -\left(R_{\alpha\beta}-\frac{1}{2}Rg_{\alpha\beta}\right)$
in equation (4) in \cite{capoz1}. Here one notes that for an arbitrary number
of dimensions $D$, the correct form of the geometric fluid is given by (\ref{E1}),
and the form (\ref{eab}) is true only in the specific case: $D=4$.
This implies that the results  in \cite{capoz1} based on the obtained $\Sigma_{\alpha\beta}$
tensor in equation (4) is correct only in four dimensions.

Defining $\phi=\mathcal{F}_\mathcal{G}(R,\mathcal{G})$ and $\psi=\mathcal{F}_R(R,\mathcal{G})$,
we have
\begin{eqnarray}
&&\nabla_\alpha \nabla_\beta \mathcal{F}_\mathcal{G}(R,\mathcal{G})=-H\dot\phi g_{\alpha\beta}+\left(\ddot\phi-H\dot\phi\right)u_\alpha
u_\beta,\nonumber\\
&&\nabla_\alpha \nabla_\beta \mathcal{F}_R(R,\mathcal{G})=-H\dot\psi g_{\alpha\beta}+\left(\ddot\psi-H\dot\psi\right)u_\alpha
u_\beta,
\end{eqnarray}
where the dot sign represents the derivative with respect to the
time coordinate $t$.
Then, we can show that $\mathcal{E}_{\alpha\beta}$  tensor in (\ref{E1}) takes the perfect fluid form (\ref{E})
 in which $A$ and $B$ read as
\begin{eqnarray}
A&=& -\frac{1}{2}\mathcal{F}(R,\mathcal{G})+\left( (D-1)\rho_1-\rho_2\right)\psi-(D-2)H\dot\psi-\ddot\psi\nonumber\\
&&+\left[\frac{1}{2}\mathcal{G}+2\rho_1(D-2)(D-3)(D-4)\left( \rho_2-\frac{D-1}{4}\rho_1\right)  \right]\phi \nonumber\\
 && -2(D-2)(D-3)\left[\rho_1(D-2)-2\rho_2 \right]H\dot\phi-2(D-2)(D-3)\rho_1 \ddot\phi,\nonumber\\
B&=&(D-2)\rho_2\psi+H\dot\psi-\ddot\psi\nonumber\\
&&-2(D-2)(D-3)(D-4)\rho_1\rho_2\phi  +2(D-2)(D-3) \left(\rho_1+2\rho_2\right) H\dot\phi\nonumber\\
&&-2\left[(D-2)(D-3)\rho_1-4(D-1)\rho_2 \right]\ddot \phi.
\end{eqnarray}
Then for any generic $\mathcal{F}(R,\mathcal{G})$  gravity theory in $D$-dimensions we have the following Proposition.

\vspace{0.5cm}
\noindent
{\bf Proposition 9:} \textit{The field equations of the general $\mathcal{F}(R,\mathcal{G})$ gravity theory are of the perfect fluid type with the energy density $\rho$ and pressure $p$ given by
\begin{eqnarray}
\rho&=&\frac{1}{\kappa}\left[\frac{(D-1)\left(D-2\right)}{2}\rho_1 -\Lambda \right]\nonumber\\
&&+\frac{1}{2}\alpha \mathcal{F}(R,\mathcal{G})+ (D-1)\left(\rho_2-\rho_1\right)\alpha\psi+(D-1)\alpha
H\dot\psi\nonumber\\
&&-\left[\frac{1}{2}\mathcal{G}+2\rho_1(D-2)(D-3)(D-4)\left( 2\rho_2-\frac{D-1}{4}\rho_1\right)  \right]\alpha\phi \nonumber\\
 && +2\rho_1(D-1)(D-2)(D-3)\alpha
 H\dot\phi+8\rho_2(D-1) \alpha\ddot\phi,\\
p&=&\frac{1}{\kappa}\left[\left(D-2\right)\left(\rho_2-\frac{1}{2}(D-1)\rho_1 \right)+\Lambda \right]\nonumber\\
&&-\frac{1}{2}\alpha \mathcal{F}(R,\mathcal{G})+\left( (D-1)\rho_1-\rho_2\right)\alpha\psi-(D-2)\alpha
H\dot\psi-\alpha\ddot\psi\nonumber\\
&&+\left[\frac{1}{2}\mathcal{G}+2\rho_1(D-2)(D-3)(D-4)\left( \rho_2-\frac{D-1}{4}\rho_1\right)  \right]\alpha\phi \nonumber\\
 && -2(D-2)(D-3)\left[\rho_1(D-2)-2\rho_2 \right]\alpha H\dot\phi-2(D-2)(D-3)\rho_1\alpha
  \ddot\phi.
\end{eqnarray}}
For $D=4$ this proposition is proved in  \cite{capoz1}. However,
as mentioned before, one notes that the proof in \cite{capoz1} is correct only for $D=4$ due to the identically vanishing property of $H_{\alpha\beta}$ in four dimensions. For cosmological applications of $\mathcal{F}(R, \mathcal{G})$
theory, see for example \cite{odin4}.
%%%%%%%%%%%%%%%%%%%%%%%%%%%%%%%%%%%%%%%%%%%%%%%%%%%%%%%%%%%%%%%%%%%%%%%%
\section{Conclusion}
In this work considering the FLRW spacetimes we have shown that the contribution of any generic modified gravity theories to the field equations is of the
perfect fluid type. As examples, we have studied the field equations of general
$\mathcal{F}(R,\mathcal{G})$ and Lovelock theories. In a forthcoming publication we investigate exact solutions of these equations by assuming certain equations of state.
%%%%%%%%%%%%%%%%%%%%%%%%%%%%%%%%%%%%%%%%%%%%%%%%%%%%%%%%%%%%%%%%%%%%%%%%
%%%%%%%%%%%%%%%%%%%%%%%%%%%%%%%%%%%%%%%%%%%%%%%%%%%%%%%%%%%%%%%%%%%%

\end{document}